# Enhanced Automated Code Vulnerability Repair using Large Language Models

David de-Fitero-Dominguez[1], Eva Garcia-Lopez[1], Antonio Garcia-Cabot[1], Jose-Javier Martinez-Herraiz[1]

[1]Universidad de Alcalá, Departamento de Ciencias de la Computación, Madrid (Spain)

david.fitero@edu.uah.es, eva.garcial@uah.es, a.garciac@uah.es, josej.martinez@uah.es

## Abstract

This research addresses the complex challenge of automated repair of code vulnerabilities, vital for enhancing digital security in an increasingly technology-driven world. The study introduces a novel and efficient format for the representation of code modification, using advanced Large Language Models (LLMs) such as Code Llama and Mistral. These models, fine-tuned on datasets featuring C/C++ code vulnerabilities, significantly improve the accuracy and adaptability of automated code repair techniques. A key finding is the enhanced repair accuracy of these models when compared to previous methods such as VulRepair, which underscores their practical utility and efficiency. The research also offers a critical assessment of current evaluation metrics, such as "Perfect Predictions", and their limitations in reflecting the true capabilities of automated repair models in real-world scenarios. Following this, it underscores the importance of using test datasets devoid of train samples, emphasizing the need for dataset integrity to enhance the effectiveness of LLMs in code repair tasks. The significance of this work is its contribution to digital security, setting new standards for automated code vulnerability repair and paving the way for future advancements in the fields of cybersecurity and artificial intelligence. The study does not only highlight the potential of LLMs in enhancing code security but also fosters further exploration and research in these crucial areas.

## Introduction

In the current information age, where digitization and software usage are prevalent in everyday life, computer security has become a primary concern in diverse fields, including the military, social and economic sectors (Ji et al., 2018). Computer systems, ranging from simple mobile applications to intricate critical infrastructure systems, are built on countless lines of code. Each line of code is vulnerable to exploitation by malicious actors, which could lead to potential financial losses, service disruptions, theft of sensitive information, reputational damage, and other harmful effects on the system (Farahbod et al., 2020).

With the exponential growth of software in all sectors of society, the impact of these vulnerabilities is becoming increasingly severe. This issue is further exacerbated by the rise of open-source software (*Octoverse 2022*, 2022), which allows anyone to view and contribute to the source code. While this development approach has several advantages, including transparency and collaboration, it may also increase the exposure of security vulnerabilities to a broader audience, thus escalating the risk of exploitation.

In addition, the increasing complexity of systems makes it increasingly difficult for humans to manually inspect code for vulnerabilities (O'Driscoll, 2022). Despite the existence of static code analysis tools and bug bounty programmes, the detection and correction of vulnerabilities remains a significant challenge (Lomio et al., 2022). This is where artificial intelligence, particularly natural language processing models, can play a critical role.

In recent years, considerable progress has been made in the field of Large Language Models (LLMs), due to their scalability and the application of the Transformer architecture (Vaswani et al., 2017). Notably, these LLMs have demonstrated revolutionary capabilities in a variety of domains, including but not limited to reasoning, mathematics, science, and language processing (Brown et al., 2020; Chowdhery et al., 2023; Du et al., 2022; Howard & Ruder, 2018; OpenAI, 2023; Rae et al., 2022; Tay et al., 2023). These leaps are due to the increasing size of models and the substantial amounts of data on which they are trained (Brown et al., 2020; Hoffmann et al., 2022; Rae et al., 2022).

The leading role of the Transformer architecture, which in its original design consisted of an encoder and a decoder (Vaswani et al., 2017), cannot be overstated. These components, used independently or in combination, form the basis of models such as BERT (Devlin et al., 2019) and T5 (Raffel et al., 2020), which use only encoder and encoder-decoder architectures respectively and employ a pre-training goal based on masked language modelling (MLM). In contrast, models such as those in the GPT family (Brown et al., 2020; OpenAI, 2023; Radford et al., 2018, 2019), leverage decoder-only architectures, employing a pre-training objective based on causal language modeling (CLM) and, more recently, new forms of pre-training such as causal infilling prediction (Rozière et al., 2023).

LLMs trained on massive text datasets have been shown to have the ability to perform novel tasks using only textual instructions or a few examples (Brown et al., 2020). This ability, known as few-shot learning, emerged as the scale of these models reached a significant size (Kaplan et al., 2020). An illustration of the potential of these models can be found in the GPT-3 model (Brown et al., 2020). This model showed that massive autoregressive (decoder-only) language models can be effectively used for few-shot learning. In such an approach, the model is only provided with a description of the task in natural language, along with a few examples of task completion. This seminal discovery triggered the development of several large autoregressive language models, including GLaM (Du et al., 2022), Gopher (Rae et al., 2022), Chinchilla (Hoffmann et al., 2022), Megatron-Turing NLG (Smith et al., 2022), and LaMDA (Thoppilan et al., 2022). These models have been instrumental in pushing the technological frontier in this field.

However, recent studies such as Hoffmann et al. (2022) suggest that optimal performance is not necessarily achieved by the largest models, but rather by models trained on more data, which suggest that models trained on extensive data can outperform larger models. In this context, the preferred model, given a target performance level, is not the fastest to train, but the fastest to infer. This insight leads to the conclusion that, while larger models may reach the desired performance levels more quickly during training, smaller models may be more efficient and cost-effective for inference when trained over a longer period of time. (Hoffmann et al., 2022).

These advances and discoveries in the field of language models provide a compelling motivation to explore their application in automated vulnerability repair. LLMs, with their ability to understand and generate language, provide a solid foundation for this task. Moreover, training these models on large and diverse datasets could enable them to learn language patterns useful for vulnerability identification and repair. The motivation for this work lies at the intersection of these challenges and opportunities. If we can train a language model based on the Transformers architecture using datasets of code vulnerabilities, we could create a tool capable of automatic repair. This tool could not only help mitigate the growing threat of security vulnerabilities, but also free human developers from the tedious and challenging task of fixing these bugs.

Advances in automated program repair (APR) have recently shifted toward the use of LLMs to address code vulnerabilities. Traditional APR approaches, including heuristic-based, constraint-based, and pattern-based techniques (Zhang et al., 2022, 2023), have been complemented by LLM-based methods that show promise in outperforming existing techniques across different programming languages (Xia et al., 2023). In particular, studies have explored LLMs for Java vulnerability repair (Wu et al., 2023) and introduced innovative systems such as SecRepair (Islam et al., 2024), which not only fixes vulnerabilities but also provides explanatory comments. In the area of C/C++ vulnerability repair, recent works such as VRepair (Z. Chen et al., 2022), VulRepair (Fu et al., 2022), and VulMaster (Zhou, Kim, et al., 2024) have made significant progress, each introducing novel approaches to improve the accuracy and efficiency of vulnerability repair.

The paper presents three primary contributions. Firstly, an innovative method for code modification representation is introduced, which represents a significant departure from previous approaches. This method simplifies changes across diverse code vulnerability types, enhancing modification flexibility and precision for more nuanced and context-aware adjustments in the repair process. Secondly, our use of recent advanced LLMs, namely Code Llama (Rozière et al., 2023) and Mistral (A. Q. Jiang et al., 2023), is a critical aspect of our approach. These models, through fine-tuning on specialized datasets rich in C/C++ code vulnerabilities, greatly enhance the accuracy and effectiveness of repairing code vulnerabilities. This highlights the potential of LLMs in this field. Finally, we also address a crucial methodological challenge in this field: the problem of data overlap between training and testing datasets. To ensure an accurate and reliable assessment of the performance of our models, we carefully curate distinct and clean datasets for training and evaluation. This approach prevents common issues such as overfitting and biased performance metrics, leading to a more robust evaluation of automated repair systems.

## Background

Vulnerability repair methods can be divided into two primary categories (Gao et al., 2022): general-purpose repair and security-oriented repair. General-purpose approaches, while not explicitly designed to repair this type of flaw, can be used to repair security vulnerabilities. For example, Angelix (Mechtaev et al., 2016) and Concolic Program Repair (CPR) (Shariffdeen et al., 2021) have proven to be effective on this task. These methods can be valuable whenever given specifications, like test cases, are sufficient to deduce the expected behavior, that is, the correct and safe functioning of the software as defined in the design requirements and validated by test cases.

On the other hand, specific-purpose methods focus on security vulnerabilities in source code. These approaches consider general security properties and focus on generating appropriate repair specifications from the identified vulnerability. A challenge in this process is that there is usually only one failed test, the exploit, which serves as input for the repair (Gao et al., 2022). This test case, which highlights the vulnerability, is often the only direct evidence and reference for starting the repair process. Unlike other scenarios where multiple test cases provide a broader context for identifying and addressing issues, here the repair mechanism must rely mainly on this single instance of failure. In the following, these two approaches are discussed, with the specific techniques used and the challenges associated with each being examined.

General-purpose repair approaches, as previously mentioned, refer to APR techniques that are not explicitly designed for security repair, yet can be applied to it (Gao et al., 2022). These approaches are based on the concept that, given enough specifications, such as test cases and assertions, the expected behavior can be inferred, and appropriate program transformations can be generated. These general-purpose methods can be divided into two strategies: traditional approaches based on search algorithms and constraints, and modern strategies that incorporate machine learning techniques.

Traditional APR approaches can be divided into three main categories (Zhang et al., 2022, 2023): heuristic-based repair techniques, constraint-based repair techniques, and pattern-based repair techniques.

Heuristic-based repair techniques employ heuristic strategies such as genetic algorithms to explore a search space built from previous patches and formulate valid patches (Le Goues et al., 2012; Martinez & Monperrus, 2016; Yuan & Banzhaf, 2020). SimFix (J. Jiang et al., 2018) is an example of this approach, which leverages error localization, donor code snippets (potentially useful code snippets for repairing a defective code section), and variable mapping, and uses similarity heuristics to adapt and select relevant donor codes, prioritizing valid patches based on multiple criteria.

Constraint-based repair techniques, on the other hand, approach APR as a constraint-solving task, using Satisfiability Modulo Theories (SMT) solvers to find feasible solutions (Durieux & Monperrus, 2016; Martinez & Monperrus, 2018; Mechtaev et al., 2016). Nopol (Xuan et al., 2017) exemplifies this approach, by comprising three phases to identify repair locations, gather execution context, and generate patches translating solutions to the SMT into source code patches.

Pattern-based repair techniques manually design repair patterns and apply them to erroneous code snippets (Koyuncu et al., 2020; Liu et al., 2019). For example, TBar (Liu et al., 2019) uses recurrent repair patterns, identifies suspicious defect location, selects appropriate repair patterns, generate patches, and test them until a suitable one is found, or all locations have been examined.

Machine learning techniques in APR focus on several stages of processing. Specifically, seven stages can be identified (Zhang et al., 2023), starting with error localization, where suspicious code elements are identified, and ending with patch correction, an additional stage to filter out overfitted patches after their validation to improve their quality. Understanding these stages is important, because although they are not strictly mandatory, they are essential to understand some of the techniques or mechanisms of different state-of-the-art approaches.

When examining state-of-the-art methods for automatically repairing software vulnerabilities, it is important to consider specific-purpose approaches that are designed exclusively to identify and fix security flaws. Various strategies and methodologies arise. Sidiroglou & Keromytis (2005) introduced a proactive approach that emphasizes early detection and correction of potentially vulnerable code. This is accomplished by employing a secure "sandbox" environment to evaluate questionable requests and verify patches. This methodology, centered on transferring troublesome buffers to dynamic memory (heap), facilitates prompt, decentralized responses and minimizes delays for all-encompassing solutions.

AutoPaG (Z. Lin et al., 2007) focuses on accelerating the generation of software patches, with an emphasis on real-time detection of out-of-bound vulnerabilities. It uses dataflow analysis to identify and fix vulnerable source-level program statements, dynamically adjusts vulnerable buffer boundaries, and compiles the patched code with unaffected code to generate a new, secure executable. In contrast, Senx (Z. Huang et al., 2019) leverages precise, human-defined security properties to address a variety of vulnerabilities, including buffer overflows and bad casts. Unlike AutoPaG, Senx is not limited to vulnerabilities within specific functions and can generate patches that span distinct functions. While both AutoPaG and Senx aim to accelerate the generation of secure software patches, Senx offers a more comprehensive and versatile approach by leveraging precise security properties and efficient patch placement to overcome the limitations inherent in the AutoPaG methodology.

VuRLE (Ma et al., 2017) represents a more modern automated tool that operates in learning and repair stages. It analyzes vulnerable codes and generates repair templates through learning. It accurately identifies and rectifies vulnerabilities in incoming codes, testing each template and applying transformative edits. VuRLE distinguishes itself with its adaptable repair capabilities and ability to learn from examples, setting it apart from earlier solutions that were more limited or required predefined patterns.

Harer et al. (2018) is a notable work that focuses on the repair of code vulnerabilities through the implementation of Generative Adversarial Networks (GANs), a subclass of artificial intelligence algorithms. This approach exploits the ability of GANs to operate on unpaired data, which is advantageous because it eliminates the need for paired examples, such as vulnerable code next to its fixed counterpart. Here, the generator uses a Neural Machine Translation (NMT) system to produce corrected versions of the input code, while the discriminator attempts to distinguish between the generator's outputs and actual instances of non-vulnerable code.

Compared to more modern approaches such as VuRLE or the GAN-based method, older techniques such as AutoPaG and Senx have a narrower focus, primarily targeting common vulnerabilities, such as buffer overflows, with high precision but limited scope. While they demonstrate efficiency and accuracy in addressing the specific vulnerabilities they were designed to address, their specialized nature may miss other types of bugs or vulnerabilities. This contrasts with the broader scope and adaptability of modern methods, which can learn and generalize to effectively address a wider range of vulnerabilities. This capability potentially provides more holistic and sustainable solutions in the ever-evolving landscape of software security.

In recent years, the application of LLMs to APR and vulnerability fixing has grown significantly. Xia et al. (2023) conducted a comprehensive study evaluating 9 state-of-the-art LLMs on APR tasks across 5 datasets in Java, Python, and JavaScript. They demonstrated that LLMs can outperform existing APR techniques, highlighting the importance of considering the code context after the buggy line and revealing a clear scaling effect where larger models generally perform better.

Building on this foundation, Wu et al. (2023) compared various LLMs (including Codex (M. Chen et al., 2021), CodeGen (Nijkamp, Pang, et al., 2023), CodeT5 (Wang et al., 2021), PLBART (Ahmad et al., 2021), and InCoder (Fried et al., 2023)) for Java vulnerability repair using two real-world benchmarks. While existing models fixed relatively few vulnerabilities, they found that fine-tuning with general APR data significantly improved performance. K. Huang et al. (2023) further explored the fine-tuning paradigm, focusing on 5 popular LLMs (CodeBERT (Feng et al., 2020), GraphCodeBERT (Guo et al., 2021), PLBART, CodeT5, and UniXcoder (Guo et al., 2022)) across Java, C/C++, and JavaScript. Their work provided insights into selecting appropriate strategies for LLM-based APR, including the impact of different code abstractions and evaluation metrics.

Further advancing the application of LLMs to APR, recent studies have explored innovative approaches to enhance the effectiveness and adaptability of these models. Ishizue et al. (2024) proposed a novel method that combines Refactory (Y. Hu et al., 2019), an APR tool, with GPT-3.5 (Ouyang et al., 2022), to improve program repair in educational contexts. Their method specifically addresses two limitations of Refactory: its inability to fix code with syntax errors and its difficulty in repairing code when there are few correct submissions available for reference. Complementing this work, Zirak & Hemmati (2024) addressed the issue of domain shift in APR, developing a domain adaptation framework that significantly improved the performance of models like TFix (Berabi et al., 2021) and CodeXGLUE (Lu et al., 2021) on unfamiliar projects.

Taking a more specialized approach, Islam et al. (2024) introduced SecRepair, an innovative system leveraging the CodeGen2 (Nijkamp, Hayashi, et al., 2023) model for vulnerability analysis and repair. SecRepair not only identifies and fixes vulnerabilities but also provides explanatory comments using a reinforcement learning paradigm augmented by a semantic reward mechanism. Tested on 6 Open Source IoT Operating Systems, it demonstrated the ability to address both zero-day and N-day vulnerabilities, reflecting the evolution towards more sophisticated and practical solutions in LLM-based code repair.

Finally, we base our state-of-the-art discussion on three key works: VRepair (Z. Chen et al., 2022), VulRepair (Fu et al., 2022) and VulMaster (Zhou, Kim, et al., 2024). VRepair adopts a variant approach that uses transfer learning to address vulnerability repair, applying knowledge from large datasets on general error correction. Therefore, VRepair is first trained on a large dataset of bug fixes, and then fine-tuned on a dataset of vulnerability fixes that is an order of magnitude smaller. This approach, using the Transformer architecture (Vaswani et al., 2017), allows for a refined context-based representation of code patches, enabling token-level modifications.

VulRepair, which addresses the limitations of VRepair, integrates the pre-trained CodeT5 component (Wang et al., 2021), which uses the T5 architecture (Raffel et al., 2020) with Byte Pair Encoding (BPE) tokenization to address out-of-vocabulary challenges. Specifically, VulRepair is fine-tuned on datasets targeting C/C++ code vulnerabilities, namely CVEFixes (Bhandari et al., 2021) and Big-Vul (Fan et al., 2020). For patch generation, it uses a beam search-based decoding method (as well as VRepair), outperforming benchmarks with a 44% success rate in "Perfect Predictions". In particular, the pre-training component (CodeT5) plays a critical role, contributing to 14% of "Perfect Predictions".

VulMaster, on the other hand, brings several key improvements. It handles longer vulnerable code, considers the structure of the code, and uses expert knowledge from the CWE system.

Another innovation is that VulMaster combines two Large Language Models, CodeT5 (Wang et al., 2021) and ChatGPT, to cover all aspects of vulnerability repair. CodeT5 is fine-tuned for this task, while ChatGPT provides additional relevant inputs. This approach leads to better performance, as shown by higher Exact Match (EM), BLEU, and CodeBLEU scores, making VulMaster more effective than previous methods.

Expanding on the highlighted studies, this work will delve into assessing the capabilities of two advanced LLMs, such as Code Llama (Rozière et al., 2023) and Mistral (A. Q. Jiang et al., 2023), in automating the repair of source code vulnerabilities in C/C++ language. These models, enriched with extensive training on various codebases, will be used to explore innovative solutions for fixing code vulnerabilities. The goal is to conduct an evaluation that shows how state-of-the-art code LLMs can adapt and be effective when dealing with code vulnerabilities.

## Material and methods

The current study focuses on using LLMs, specifically Code Llama (Rozière et al., 2023) and Mistral (A. Q. Jiang et al., 2023), for a specialized task: fixing software vulnerabilities in C/C++ code. This task requires the model to effectively and efficiently correct identified issues. The proposed method involves fine-tuning these modern LLMs on datasets specifically containing C/C++ code vulnerabilities, namely Big-Vul (Fan et al., 2020) and CVEFixes (Bhandari et al., 2021). The decision to fine-tune, rather than employ prompt engineering, or other non-tuning strategies, arises from the need for efficient code change representation and minimized token usage in code generation. This approach leverages the models' pre-existing programming knowledge, particularly in C, while increasing their focus on vulnerability-specific patterns.

We will use the same dataset and data splits as VulRepair (Fu et al., 2022), leveraging the work done by VRepair authors (Z. Chen et al., 2022) on CVEFixes (Bhandari et al., 2021) and Big-Vul (Fan et al., 2020) datasets to ensure comparable results. We will evaluate our approach against VulRepair, VRepair, and VulMaster (Zhou, Kim, et al., 2024) to provide a comprehensive evaluation of our work within the current landscape of automated vulnerability repair. Our fine-tuning strategy incorporates innovative techniques such as Low Rank Adaptation (LoRA) (E. J. Hu et al., 2021), which optimizes training by updating only a small set of model parameters, and 4-bit quantization (Dettmers & Zettlemoyer, 2023), which improves training efficiency by compressing model weights to save memory. Throughout our analysis, we detail the data format used and outline our performance evaluation methods, providing a clear picture of the ability of our fine-tuned models to address C/C++ code vulnerabilities.

### Data definition

The dataset is crucial, and its definition requires careful consideration. As stated before, we will employ a combination of Big-Vul (Fan et al., 2020) and CVEFixes (Bhandari et al., 2021) as the dataset in the studies conducted by VulRepair (Fu et al., 2022), VRepair (Z. Chen et al., 2022) and VulMaster (Zhou, Kim, et al., 2024). We chose this dataset combination due to its comprehensive collection of C/C++ code vulnerabilities, which offers a varied and abundant pool of samples necessary for successful model training. This procedure also adheres to established benchmarks in the field, ensuring a more significant foundation for comparing results.

These datasets (Big-Vul and CVEFixes) were created through similar methodologies but by independent research teams. Both datasets were compiled by crawling CVE (Common Vulnerabilities and Exposures) databases to extract vulnerability information such as CWE (Common Weakness Enumeration) and CVE IDs. The researchers then developed project-specific crawlers to obtain git commit links that fix these vulnerabilities. Big-Vul contains 3,754 vulnerabilities from 348 projects, categorized into 91 CWE IDs, spanning from 2002 to 2019. CVEfixes, slightly larger and more recent, includes 5,365 vulnerabilities from 1,754 projects, covering 180 CWE IDs, with data from 1999 to 2021.

We rely on the dataset published on the Hugging Face platform[1] by the authors of VulRepair (Fu et al., 2022), which is a combination of the aforementioned datasets. We carefully replicated their test set to ensure comparability with previous research. However, our analysis revealed a significant overlap between the training and test sets, with about 40% of test samples also present in the training data. This overlap may lead to overfitting (a situation where the model performs exceptionally well on familiar data but struggles with new, unseen vulnerabilities) and thus presents a challenge in accurately assessing model performance.

For that reason, we use two different training subsets to thoroughly evaluate the performance of our model and avoid overfitting. The first training set tries to replicate the exact same samples as the ones used in VulRepair (Fu et al., 2022) and VRepair (Z. Chen et al., 2022), serving as a baseline for comparison. The second is a cleaned version with no overlap between training and test data, simulating real-world conditions where the model encounters entirely new vulnerabilities. Notably, the VulMaster study (Zhou, Kim, et al., 2024) also acknowledges this overlap issue and creates a new dataset based on CVEFixes and Big-Vul without this overlap. By training on both our sets and testing on the same test set, we can compare results and understand how data overlap affects performance metrics. Furthermore, the VulMaster study will allow us to validate our findings and provide additional insights into the impact of data cleaning on model performance across different approaches. This comprehensive approach enables a more thorough and reliable evaluation of the capabilities of our models in real-world vulnerability repair scenarios.

Table 1 displays the dataset composition used in this study. The first row lists the replication of the original dataset, which comprises a training set of 6,429 samples, a validation set constituting 5% of the training set size with 338 samples, and a test set of 1,706 samples. The second row shows the refined dataset, where the training set has been reduced to 4,163 samples and the validation set to 219 samples after removing overlaps and duplicates. The test set size remains unchanged as it is the same for both datasets.

| Dataset | Train | Validation | Test |
| --- | --- | --- | --- |
| CVEFixes+Big-Vul (original) | 6,429 | 338 | 1,706 |
| CVEFixes+Big-Vul (refined) | 4,163 | 219 | 1,706 |

*Table 1. Final datasets split size.*

Figure 1 shows the dataset configuration and splits used in our vulnerability repair study, illustrating the flow from the original Big-Vul and CVEFixes datasets to the final training, validation, and test sets for both the original and refined datasets.

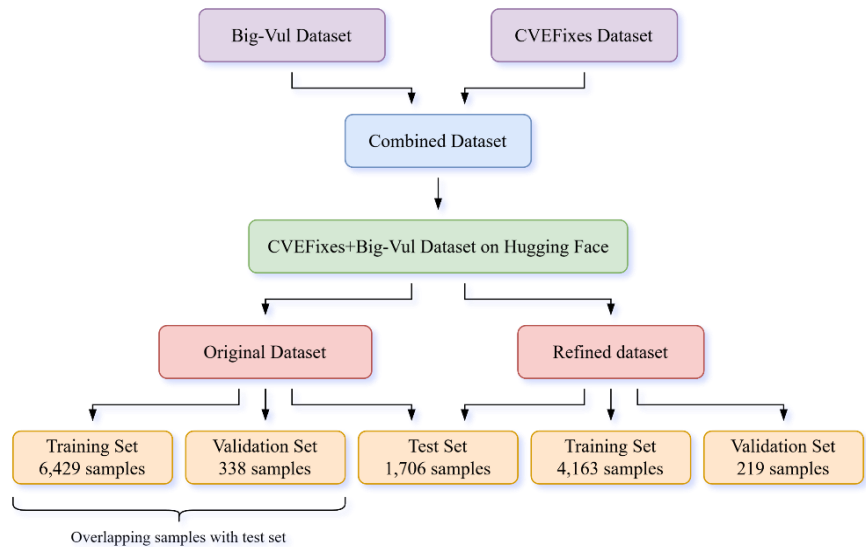

*Figure 1. Dataset Configuration and Splits*

## Data format

The data format used in this project is carefully structured to allow efficient processing and output generation by the LLMs. Figure 2 shows the format.

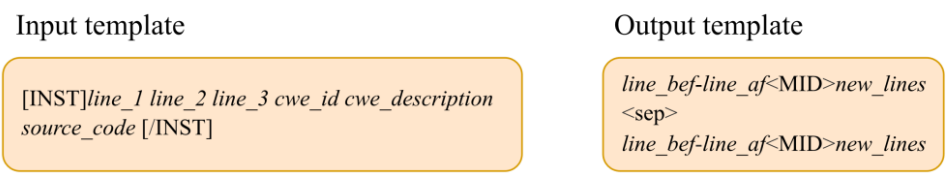

*Figure 2. Input / output template.*

In this format, "[INST]" and "[/INST]" serve as special marking tokens that indicate the beginning and end of the input prompt to the model. These tokens are based on the tokens used by instruction-tuned models fine-tuned over Llama 2 (Touvron et al., 2023). Within the input prompt, "line_1", "line_2", and "line_3" represent the line numbers in the source code identified as vulnerable code. The inclusion of these specific lines helps in focusing the attention of the model on the parts of the code that require modification. Following this, "cwe_id" and "cwe_description" provide context about the type of vulnerability, based on the Common Weakness Enumeration (CWE) system[2]. This additional information may further aid the model in understanding the nature of the vulnerability. Finally, the "source_code" section, annotated with line numbers starting from 0, provides the complete context in which the vulnerable lines reside.

The output format of the model in this project is finely tuned to provide specific and practical suggestions for code changes. The indicators "line_bef" and "line_af" respectively mark the line before and the line after where the new code will be inserted in the source code. The "<MID>" sequence following these indicators presents the new lines of code suggested by the model to be placed between "line_bef" and "line_af". Figure 3 provides an illustrative example of a processed sample of the training dataset split.

---

[2] https://cwe.mitre.org/

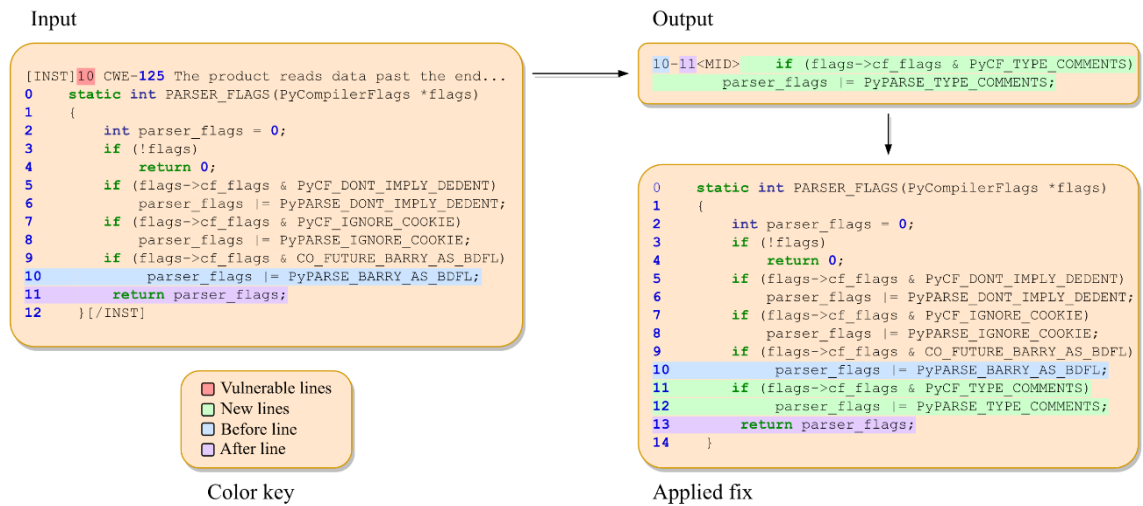

*Figure 3. Example of a processed sample of the train split.*

The nature of the change depends on the context. If the model proposes a change such as "X-Y<MID>", where X and Y are consecutive line numbers (as illustrated in Figure 3), the new lines after the "<MID>" token should be placed between those line numbers and would represent an addition. If X and Y are non-sequential, then every line between X and Y should be replaced with the new lines after the "<MID>" token, denoting a replacement modification. If the model suggests deleting every line between X and Y, the area after the "<MID>" tag would be left blank. Separator "<sep>" is used when suggesting modifications at multiple code locations, ensuring each fix is distinct and can be individually addressed.

The format chosen for data presentation in this study offers several distinct advantages that enhance the effectiveness and efficiency of the LLMs used to correct code vulnerabilities. One significant advantage of this format is its versatility in suggesting any type of code modifications. This format allows the model to suggest insertions, replacements, and deletions with little or no redundant lines of code. Another significant advantage of this format is the clarity it provides to the output of the model. The use of the "<MID>" delimiter marks the beginning of the modification suggestion, making it easy to identify and understand the proposed changes. The structured nature of the output not only reduces the risk of misinterpretation but also simplifies the application of these changes to the original code. This clarity and ease of implementation are crucial for efficiently using the outputs of the model in real coding scenarios.

Additionally, the selected format notably decreases redundancy in the generated output of the model. By focusing only on generating the new lines of code to be inserted between "line_bef" and "line_af", the model does not need to reproduce any part of the existing source code that remains unchanged. This tactic eliminates needless repetition and streamlines the output to include only the necessary adjustments. The reduction in redundancy also improves the inference speed of the model. With fewer tokens to generate and only the necessary modifications focused on, it reduces the computational load, resulting in faster processing times. This efficiency proves to be incredibly valuable in scenarios where quick response times are crucial.

Our methodology also addresses a significant limitation observed in previous studies. In prior approaches, a minimal set of tokens, extracted from Clang tokenizer, provided the context for code modification, relying on a constrained set of three context tokens. However, this approach had a weakness: it could potentially misalign suggested changes with their intended location in the source code. Specifically, the context tokens could appear earlier in the source code than the actual location of the required change, leading to erroneous changes and complicating the implementation process.

The proposed format mitigates this problem by ensuring that any change suggested by the model is straightforward to implement, with a greatly reduced chance that the change will be misplaced. By clearly delimiting the lines of code before and after the intended change (indicated by "line_bef" and "line_af"), and by specifying the exact changes using the "<MID>" delimiter, the format precisely guides the application of changes. This structure effectively eliminates the risk of misplaced modifications inherent in the previous token-based context approach. As a result, it simplifies the process of translating the suggestions of the model into actual code, thereby increasing the accuracy and efficiency of the code correction process.

### Fine-tuning process

In our fine-tuning process, we employ QLORA (Quantized Low Rank Adaptation) (Dettmers et al., 2023), an advanced technique that allows efficient training of large language models. QLORA works by using a 4-bit quantized version of the pre-trained model and applying LoRA (E. J. Hu et al., 2021), which injects trainable low-rank decomposition matrices into the linear layer of the models while freezing the pre-trained model weights. This significantly reduces the amount of memory used, while barely affecting performance. QLORA also introduces a new 4-bit data type called NormalFloat, optimized for neural network weights, and uses double quantization and paged optimizers to further reduce memory requirements. These innovations enable fine-tuning of much larger models than previously possible with standard methods.

We also incorporate NEFTune (Noisy Embedding Fine-Tuning) (Jain et al., 2023). NEFTune is a technique that enhances the instruction fine-tuning of language models. The process involves the integration of a random noise vector into the embedding vectors during training. This noise, introduced at each step of training, is precisely calibrated, sampled uniformly, and scaled according to the sequence length, embedding dimension, and a tunable parameter α. The introduction of such noise aims to regularize the training process, preventing the model from overfitting to the training data and encouraging it to generalize better to new, unseen data.

Our hardware consists of a system containing four NVIDIA Quadro RTX 5000 GPUs, each with a 16GB VRAM capacity. This setup allows us to work with models up to 7 billion parameters in size. We selected this parameter limit with dual considerations in mind: firstly, to adhere to the VRAM capacity constraints; secondly, to optimize inference speed. Using models containing up to 7 billion parameters, we can maintain a balance between computational depth and operational efficiency. Larger models have the potential to be more powerful, but their increased demand for memory allocation and longer processing times exceed our hardware capabilities without providing a significant increase in performance. Maintaining a balance between computational depth and the agility required for fast inference ensures that our models are both powerful and responsive within the constraints of our hardware configuration.

| Hyperparameter | Value |
| --- | --- |
| Batch size | 1 |
| Gradient accumulation steps | 64 |
| Learning rate scheduler type | Cosine |

| Learning rate | 3e-4 |
|---|---|
| Optimizer | AdamW |
| Lora R | 8 |
| Lora alpha | 16 |
| Lora target modules | All dense layers |
| NEFTune noise alpha | 5 |
| Epochs | 8 |

*Table 2. Fine-tuning hyperparameters.*

Table 2 displays the hyperparameters selected for the fine-tuning process of our models. To address GPU memory limitations when processing big models, we combined a batch size of 1 with 64 gradient accumulation steps. We chose the cosine learning rate scheduler for its cyclical ability to adjust the learning rate. We employed the AdamW optimization algorithm (Loshchilov & Hutter, 2019), an adaptation of the standard Adam optimizer that includes weight decay regularization, with a learning rate set to 3e-4. These are typical parameters used to fine-tune LLMs (Touvron et al., 2023).

When determining the hyperparameters for LoRA (E. J. Hu et al., 2021), we chose a rank of 8 to provide a reasonable balance between computational efficiency and model performance. Selecting a higher rank could enable the capture of more complex model adaptations, but prior evaluation suggested that the incremental gains in performance did not proportionately offset the additional computational requirements. As recommended in the original LoRA research, we have fixed the alpha value for scaling the learned weights at 16. Additionally, we have expanded the LoRA adapters to include all dense layers, based on empirical evidence indicating that such a comprehensive approach can lead to improvements that are comparable to those resulting from full-parameter fine-tuning (Dettmers et al., 2023). The parameter noise alpha in NEFTune has been set to 5 based on empirical evidence (Jain et al., 2023). This value indicates a balance between introducing beneficial noise and preserving the original integrity of the data.

### Evaluation metrics

Our evaluation methodology is designed to provide a comprehensive assessment of the effectiveness of our trained models in fixing code vulnerabilities. We employ a single primary metric, "Perfect Predictions" (PP), while using different decoding techniques to explore the capabilities of our models. Additionally, we introduce efficiency metrics including total execution time, tokens generated, and generated patches per second, offering insights into the computational performance of the models.

The "Perfect Predictions" metric sets a rigorous standard for accuracy. It classifies the output of a model as correct only if it exactly matches the reference solution. When generating multiple patches per sample, we count it as a "hit" if at least one patch exactly matches the reference. This approach ensures high precision but has a limitation: it may overlook functionally equivalent solutions that differ syntactically, a common occurrence in software development. Despite this drawback, "Perfect Predictions" serve as a clear, though strict, measure of accuracy and allow for comparisons between studies in the field.

For decoding, we will use beam search with a limit of five beams, due to hardware constraints. Beam search is a commonly accepted method that systematically generates possible sequences, providing precise forecasts by examining numerous sequence hypotheses during decoding (Tunstall et al., 2022). Second, we will employ a sampling decoding technique with a temperature of 0.8, similar to the HumanEval benchmark (M. Chen et al., 2021). This approach introduces randomness, resulting in diverse outputs from the same input. We generate up to five patches per instance to maintain comparability with beam search. This evaluation approach aligns closely with established practices in recent vulnerability repair studies (Z. Chen et al., 2022; Fu et al., 2022; Zhou, Kim, et al., 2024).

This evaluation methodology, using beam search and sampling along with efficiency metrics, intends to provide a comprehensive and in-depth assessment of the predictive accuracy and reliability of our models. Through the implementation of these techniques, our goal is to validate the effectiveness of our fine-tuning efforts and provide meaningful insight into the operational strengths, limitations, and practical applicability of the models in real-world scenarios.

## Results

Figure 4 and Figure 5 show performance results for the refined and original datasets, respectively. The labels "Sampling" and "Beams" are followed by numbers indicating how many distinct solutions were generated for each test case. For example, "Sampling 5" means that five unique solutions were generated per test case, while "Beams 5" uses beam search with a width of five, also producing five outputs per test case. A higher number of samples or wider beam search increases solution diversity, potentially improving accuracy by providing more opportunities for a correct prediction.

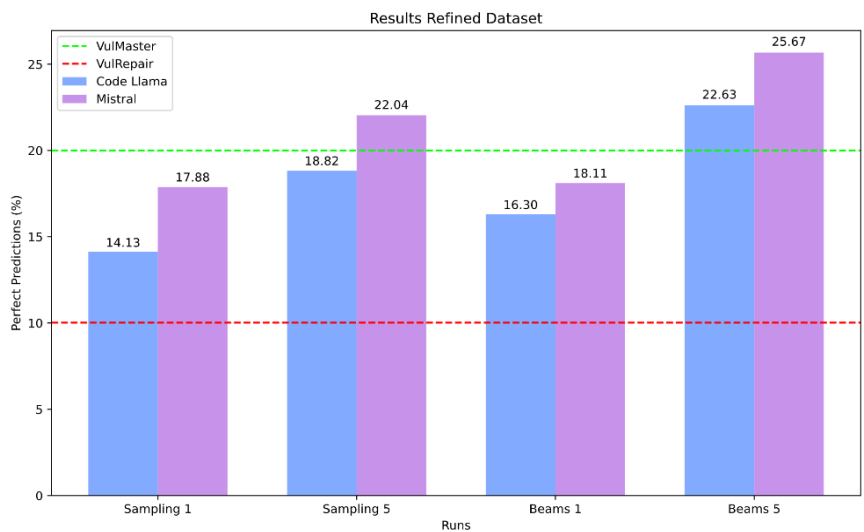

*Figure 4. Results on the refined dataset.*

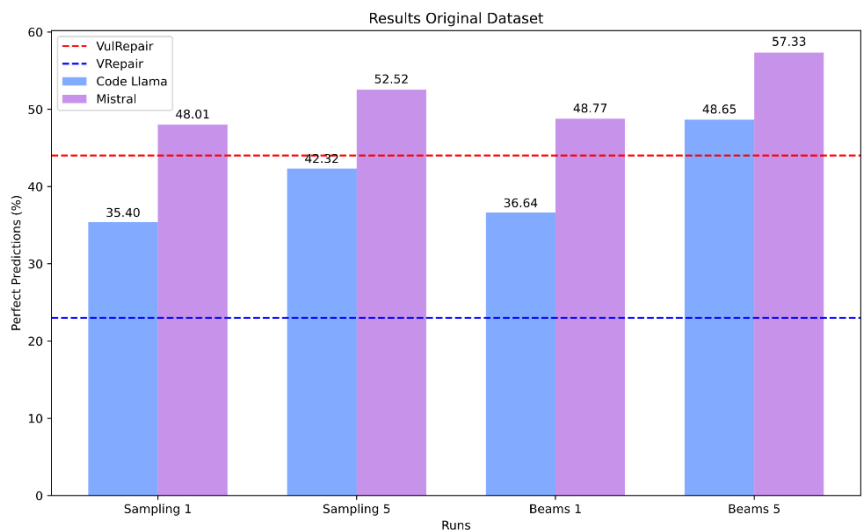

*Figure 5. Results on the original dataset.*

In the original dataset, Code Llama achieved accuracies of 35% and 42% for sampling 1 and sampling 5, respectively. These results suggest that the model accurately predicted 35% of test cases when using the first solution (sampling 1) and showed a slight improvement to 42% when expanding to the top five solutions (sampling 5). For beams 1 and 5, the accuracies are 37% and 49%, respectively, suggesting that a broader beam search width, in this case beam 5, leads to better results. In comparison, the Mistral model demonstrated higher accuracies in this dataset, achieving 48% for sampling 1 and 53% for sampling 5, and 49% for beams 1 and 57% for beams 5.

In the refined dataset, which does not include samples from the training set in the test set, the performance of both models notably decreased. The accuracy of Code Llama dropped to 14% for sampling 1 and 19% for sampling 5, and further to 16% for beams 1 and 23% for beams 5. Similarly, the accuracy of Mistral dropped to 18% for sampling 1 and 22% for sampling 5, and to 18% for beams 1 and 26% for beams 5. Despite this decrease, it is noteworthy that the top performance of our Mistral model still outperforms both VulMaster (20%) and VulRepair (16.8%) on comparable datasets. This comparison highlights the robustness of our approach, particularly Mistral, in dealing with unseen vulnerabilities, even when faced with a more challenging, non-overlapping dataset.

| Rank | CWE-Type | Name | VulRepair | VulMaster | Mistral (ours) |
|---|---|---|---|---|---|
| 1 | CWE-787 | Out-of-bounds Write | 3/58 | 12/58 | 15/53 |
| 2 | CWE-79 | Cross-site Scripting | 0/1 | 1/1 | 0/1 |
| 3 | CWE-89 | SQL Injection | 0/4 | 1/4 | 0/5 |
| 4 | CWE-416 | Use After Free | 1/60 | 6/60 | 19/55 |
| 5 | CWE-78 | OS Command Injection | 0/4 | 0/4 | 0/3 |
| 6 | CWE-20 | Improper Input Validation | 18/128 | 33/128 | 40/152 |
| 7 | CWE-125 | Out-of-bounds Read | 12/156 | 20/156 | 32/170 |
| 8 | CWE-22 | Path Traversal | 0/5 | 0/5 | 1/8 |
| 9 | CWE-352 | Cross-Site Request Forgery | 0/1 | 0/1 | 0/2 |
| 10 | CWE-434 | Dangerous File Type | - | - | - |

*Table 3. Comparison of "Perfect Predictions" for top 10 CWE across VulRepair, VulMaster, and Our Mistral Model*

The Table 3 shows the top 10 CWE, comparing the performance of three approaches: VulRepair (Fu et al., 2022), VulMaster (Zhou, Kim, et al., 2024), and our Mistral-based model. For this comparison, we used our Mistral model trained on the training set of the refined dataset and evaluated it using beam search decoding with a beam size of 5. It is important to note that the differences in the total number of samples per CWE between the models are due to our use of the publicly available Hugging Face dataset, while the VulMaster and VulRepair results are based on the dataset created in the VulMaster study, which does not contain the same test samples as the publicly available dataset. Despite these differences in data sources, the comparison provides valuable insights into the relative strengths of each approach in addressing different types of code vulnerabilities.

Our Mistral-based model demonstrates notable improvements in two key areas. For "Out-of-bounds Write" vulnerabilities (CWE-787), our model successfully fixed 15 out of 53 cases, outperforming both VulRepair (3/58) and VulMaster (12/58). Similarly, for "Use After Free" issues (CWE-416), Mistral shows significant progress with 19 fixes out of 55 cases, compared to VulRepair with 1 out of 60 and VulMaster with 6 out of 60. These results highlight the potential of our approach in addressing critical code vulnerabilities, particularly in memory management scenarios.

| Model | Run type | Total time (s) | Total tokens generated | Patches per second |
|---|---|---|---|---|
| Code Llama | Beam search 1 | 567.06 | 182,692 | 3.01 |
| Code Llama | Beam search 5 | 4,658.97 | 765,949 | 0.37 |
| Code Llama | Sampling 1 | 567.32 | 181,285 | 3.01 |
| Code Llama | Sampling 5 | 938.44 | 913,231 | 1.82 |
| Mistral | Beam search 1 | 527.28 | 194,619 | 3.24 |
| Mistral | Beam search 5 | 5,333.54 | 859,740 | 0.32 |
| Mistral | Sampling 1 | 521.55 | 189,698 | 3.27 |
| Mistral | Sampling 5 | 941.89 | 947,444 | 1.81 |

*Table 4. Performance comparison of Code Llama and Mistral models across different run types and metrics.*

Additionally, Table 4 presents efficiency metrics for Code Llama and Mistral models, derived from generating patches over the entire test set of 1,706 samples. These metrics highlight the models' performance in terms of speed and resource usage. The "Total time" column shows the overall duration for generating patches across all test samples. "Total tokens generated" indicates the output volume, reflecting the models' productivity. "Patches per second" measures throughput, with higher values indicating faster processing. Notably, these inferences were conducted using the vLLM inference engine (Kwon et al., 2023), which implements the PagedAttention algorithm for efficient key-value cache management. The inferences were performed in batches to optimize performance, using the same hardware as the one used for fine-tuning. This approach allows for improved throughput by enabling larger batch sizes through more efficient memory usage.

```
CWE-119 - The product performs operations on a memory buffer, but it can read from or
write to a memory location that is outside of the intended boundary of the buffer.

  static int vorbis_finish_frame(stb_vorbis *f, int len, int left, int right)
  {
      int prev,i,j;
      if (f->previous_length) {
          int i,j, n = f->previous_length;
          float *w = get_window(f, n);
+         if (w == NULL) return 0;
          for (i=0; i < f->channels; ++i) {
              for (j=0; j < n; ++j)
                  f->channel_buffers[i][left+j] =
                      f->channel_buffers[i][left+j]*w[ j] +
                      f->previous_window[i][ j]*w[n-1-j];
          }
      }
      prev = f->previous_length;
      f->previous_length = len - right;
      for (i=0; i < f->channels; ++i)
          for (j=0; right+j < len; ++j)
              f->previous_window[i][j] = f->channel_buffers[i][right+j];
      if (!prev)
          return 0;
      if (len < right) right = len;
      f->samples_output += right-left;
      return right - left;
  }
```

*Figure 6. Generated patch addressing a CWE-119 buffer overflow vulnerability in a sample from the test subset*

Figure 6 shows an example of a patch generated by our fine-tuned Mistral model. It exhibits a CWE-119 vulnerability, risking buffer overflow due to unchecked memory operations. To address this, our model generated a patch that adds a crucial null check for the "w" pointer returned by "get_window", effectively preventing potential crashes or undefined behavior. To create this targeted fix, the model demonstrated a comprehensive understanding of the context of the function, C error handling practices, memory safety principles, and the importance of validating function returns. This showcases the capability of the model to identify subtle vulnerabilities and propose concise, effective solutions.

## Discussion

In this section, we explore our implications with respect to the methodologies employed by previous studies, including VulRepair (Fu et al., 2022) and VRepair (Z. Chen et al., 2022). An area of dispute is the use of the original dataset by these previous works, which we found to have a significant overlap of approximately 40% between the training and test samples. The considerable overlap in the data raises concerns regarding the accuracy of the models used in these studies. This is because the significant redundancy is likely to have led to overfitting, a situation in which the performance of a model is limited to the training data, limiting its practical applicability. Even though the VulRepair and VRepair models may have impressive performance metrics, their familiarity with the test samples may give a misleading sense of accuracy. Therefore, any assertions regarding the efficiency of the models and their capability to apply to genuine, unseen vulnerabilities are weakened by this deficiency in the dataset.

Our analysis of the generalization capabilities of models trained on the original dataset supports these concerns. The refined dataset, which eliminated the overlap between the training and test sets, showed a significantly lower performance on the test set, revealing a potential overestimation of the capability of the model in previous research. For example, while Mistral achieved 57% accuracy on beams 5 in the original dataset, this value dropped to 26% in the refined dataset, which is more than half. The disparity raises doubts about the reliability of conclusions drawn from such contaminated datasets and underscores the need for rigorous dataset cleaning to ensure valid evaluations of model performance.

It is important to highlight that due to technical limitations, our results are based on a maximum beam size of 5, while the studies we are comparing against used a beam size of 50. Despite this constraint, our models demonstrate superior performance. For instance, our Mistral model achieved a "Perfect Prediction" rate of 25.67% with a beam size of 5 on the refined dataset. This surpasses the performance of VulMaster, which reported a 20.0% "Perfect Prediction" rate.

Analyzing the efficiency data alongside the "Perfect Predictions" results reveals compelling trends in model performance. Mistral demonstrates superior accuracy, particularly with beams 5 (26% vs 23% for Code Llama on the refined dataset), while also showing comparable or better efficiency in several aspects. For instance, Mistral is faster in beams 1 (527.28s vs 567.06s for Code Llama) and in sampling 1 (521.55s vs 567.32s), while generating more tokens in these configurations. However, Code Llama exhibits greater efficiency in beams 5, completing the task in less time (4,658.97s vs 5,333.54s for Mistral). These differences underscore the importance of considering both accuracy and efficiency when selecting models for vulnerability repair tasks.

A closer look at Table 4 reveals an interesting pattern in the relationship between beam search and sampling strategies. For both models, sampling 5 generates significantly more tokens than beam search 5 (913,231 vs. 765,949 for Code Llama, 947,444 vs. 859,740 for Mistral), while taking significantly less time. This suggests that sampling may be a more efficient method for generating diverse repair candidates, potentially allowing a broader exploration of the solution space without the computational overhead of beam search. This finding could be valuable for optimizing vulnerability repair systems, especially in scenarios where quick generation of multiple repair options is prioritized over the structured approach of beam search.

In real-world applications, the ability to produce high-quality solutions quickly and accurately is paramount. Our models demonstrate superior performance by producing more accurate solutions with a smaller beam size, which is particularly beneficial in real-world scenarios where efficiency and speed are critical. This approach is consistent with the principle that it is better to generate fewer but higher quality solutions than a larger number of inferior solutions. The balanced performance of both Mistral and Code Llama in different configurations highlights their suitability for different real-world applications, allowing developers to choose the most appropriate model based on their specific requirements for accuracy, speed, and resource usage in vulnerability repair tasks.

Furthermore, it is important to consider the role of quantization techniques like AWQ (Adaptive Weight Quantization) (J. Lin et al., 2023) and GPTQ (Frantar et al., 2023) in mitigating issues surrounding model size, inference speed, and memory consumption. For example, AWQ reduces the overall VRAM footprint of our models to 5.5 GB, demonstrating that even large models can be optimized for efficient deployment. These optimizations enable us to maintain the advantages of larger, powerful models, such as their improved prediction and understanding capabilities, while also ensuring manageable inference speeds and memory sizes. These improvements guarantee that the models are both theoretically strong and practically usable, aligning with the needs of modern software development and security applications.

It is also essential to examine the noticeable performance discrepancy between the two 7-billion-parameter models used in our study: Code Llama (Rozière et al., 2023) and Mistral (A. Q. Jiang et al., 2023). Although Mistral was initially expected to perform slightly worse than Code Llama on code tasks (A. Q. Jiang et al., 2023), it performed better in every evaluation setting with identical training conditions. This difference in performance provides further evidence of the superior adaptability and superior performance when compared to Llama 2 (A. Q. Jiang et al., 2023; Touvron et al., 2023).

Another factor that could potentially obscure the true effectiveness of the models is the metric used to evaluate them. The "Perfect Predictions" metric, which is used to assess the accuracy of model-generated code fixes, sets a high bar for accuracy. This metric considers a prediction to be correct only if it exactly matches the reference output. Although this standard is clear, it may not fully reflect the capabilities of the models, particularly when repairing code vulnerabilities. This rigidity overlooks the fact that there may be multiple valid solutions to a single programming problem, and only recognizes one solution as correct. As a result, the program may produce an acceptable correction that is deemed incorrect due to minor differences from the reference. This can lead to an underestimation of the model's true performance, and generating a dataset that includes all correct coding solutions is a major challenge. Thus, while "Perfect Predictions" offer a simplistic evaluation approach, it may not be aligned with the practical aspects of code repair, underscoring the need for a nuanced interpretation of model performance.

Another factor to consider when evaluating vulnerability solutions is the mismatch between the CWE identified in the source code and the reference solutions provided. The CWE is a classification system used for identifying software vulnerabilities, and each vulnerability in the source code should ideally have a corresponding, precise reference solution. However, discrepancies can occur due to errors in annotation or challenges in mapping vulnerabilities to their appropriate fixes. This task is complicated by the presence of "noise", or code modifications in the dataset that do not contribute to the resolution of the CWE, and in some cases, modifications may not fully address the CWE. Additionally, the process of automatically and accurately annotating these relationships is a significant challenge, which adds complexity to this task.

This misalignment can significantly affect performance evaluations, particularly when using metrics like "Perfect Predictions" that rely on precise matches between the output of the model and reference solutions. If the reference solutions are incorrect or do not correspond to the identified CWEs, the model may be unfairly penalized. This may happen even if it produces correct solutions that do not match the erroneous references.

Moreover, these discrepancies can result in problems during the training phase of the model. If the data for training does not accurately link CWEs to their proper solutions, the model might acquire inaccurate associations. This misalignment can hinder the capacity of the model to generalize its learning to new, unseen code, which is critical for effective vulnerability repair. A poorly matched data trained model may face challenges in real-world situations, thereby reducing its effectiveness in fixing and resolving software vulnerabilities. For example, consider a sample of the test set along with the proposed solution by the fine-tuned Mistral model in Figure 7.

*Figure 7. Representation of evaluation metric and dataset limitations*

The proposed solution in Figure 7 simply replaces the *vpx_memcpy* function with *memcpy*, which does not adequately address the CWE-119 vulnerability affecting the code. CWE-119 is concerned with operations on a memory buffer that may read from or write to a location outside the intended buffer boundary, leading to buffer overflows. In the reference code, *memcpy* is used to copy data from *cpi->mb_activity_map* to *sortlist*, with the size of the data being copied based on *cpi->common.MBs*. The primary issue here is the lack of checks to ensure that the size of *cpi->mb_activity_map* does not exceed the allocated size of *sortlist*. Without such checks, there is a risk that *memcpy* could attempt to copy more data than *sortlist* can hold, potentially causing a buffer overflow. Additionally, the use of *memcpy*, which does not inherently perform boundary checks, does not reduce this risk. Therefore, the reference solution also fails to adequately mitigate the risk of buffer overflows, which is central to addressing CWE-119.

Further to the discussion, it is noted that in Figure 7, the solution proposed by the LLM appears to be identical except for some differences in whitespace (marked in red). However, since this minor difference exists, this example is considered incorrect. This underscores the significant problem with using the perfect prediction metric, as it may underestimate the true performance of the model. This metric lacks the ability to consider nearly identical solutions that differ only in minor formatting details, despite its aim for precision. This highlights the requirement for a more nuanced method of assessing model performance, notably in situations where the correctness of a solution is not strictly binary and minor variations do not affect the functional outcome.

Practical implications of this study are significant for the field of software security and development. Our approach, which combines advanced LLMs with an innovative method for representing code modifications, shows great promise for real-world application in automated vulnerability repair. The superior performance of our models, particularly in terms of accuracy and efficiency, suggests that this technology could be integrated into existing development pipelines to provide fast, reliable fixes for common code vulnerabilities. This integration could significantly reduce the time and resources typically spent on manual code review and repair, allowing development teams to focus on more complex aspects of software creation and innovation.

Additionally, the capacity of our models to generalize to unseen vulnerabilities, as demonstrated by their performance on the refined dataset, opens possibilities for proactive security measures across various sectors. In industrial environments where software security is critical, such as industrial control systems or critical infrastructure, these models could offer an additional layer of protection. This would not only improve overall security but also reduce downtime and costs associated with manual security updates. Beyond industrial applications, companies could potentially use these models to repair vulnerabilities in legacy code bases, or to provide real-time suggestions for secure coding practices during the development process.

## Conclusions

This study represents a significant advancement in automated code vulnerability repair, achieved through the creation of a new and efficient method for coding modifications. Our methodology surpasses previous approaches such as VulRepair (Fu et al., 2022), VRepair (Z. Chen et al., 2022) and VulMaster (Zhou, Kim, et al., 2024), by providing a more versatile, efficient and accurate solution for addressing different types of code vulnerabilities.

Central to our innovative approach is the design of a representation method that simplifies the application of changes to source code, regardless of the type and number of modifications required. This method addresses the limitations of token-based approaches by greatly reducing the probability of ambiguity when applying model-proposed modifications. These token-based approaches, such as those used in VulRepair and VRepair, often struggle to precisely locate changes which can lead to errors when integrating those modifications. Our method offers a dynamic framework, which allows models to comprehend and implement adjustments in a context-aware manner, significantly increasing the applicability of automated code repair.

Two LLMs were used for our study, specifically Code Llama (Rozière et al., 2023) and Mistral (A. Q. Jiang et al., 2023). These models were fine-tuned on specialized datasets that contained numerous C/C++ code vulnerabilities, enabling us to fully use their advanced pattern recognition and learning capabilities. This process of fine-tuning was critical not just for repairing vulnerabilities, but also for evaluating the performance of our novel code modification representation method. The use of our representation method, together with advanced LLMs, has led to significant enhancements in repairing code vulnerabilities.

Addressing data duplication between training and testing datasets is another key focus of our study. Our research revealed that this data overlap problem was prevalent in previous studies, including those conducted by VulRepair (Fu et al., 2022) and VRepair (Z. Chen et al., 2022). As stated in our analysis of their methods, their models were trained and tested on datasets with a significant degree of overlap, approximately 40%. This overlap carries the risk of over-fitting the models to the data on which they were trained, leading to an unrealistic assessment of their effectiveness in real-world scenarios.

To investigate one of the central focuses of the study, a dual dataset approach was used. The first dataset (the original one) provided a basis for performance evaluation. The second dataset, a more refined dataset, was meticulously cleaned of any overlap between the training and test sets. By doing so, we were able to accurately assess the effect of data overlap on model performance. The findings of our study were quite revealing. Models trained on the original dataset displayed inflated performance metrics that noticeably decreased when tested on the refined dataset. This disparity highlights the importance of our approach, which uses separate, clean datasets for training and testing, ensuring a more realistic assessment of the generalization capabilities of the models.

A remarkable result in terms of model performance also emerged. We found that, despite a beam size of only 5, our models surpassed, by a significant margin of 13%, the VulRepair (Fu et al., 2022) results, which used a beam size of 50. These findings showcase that our models are not only more efficient but also robust in handling complex vulnerability repair tasks, a critical factor in real-world applications. Additionally, our findings revealed that Mistral (A. Q. Jiang et al., 2023) performed significantly better than Code Llama (Rozière et al., 2023) in these tasks.

Furthermore, we encountered significant limitations with the datasets and metrics used. A major problem arises from the mismatch between CWEs and their corresponding solutions in the training data. This discrepancy can cause the models to make incorrect associations, which can severely compromise its generalization capabilities. Such misalignments can significantly reduce the effectiveness of the model in accurately repairing vulnerabilities in new and unseen code, thereby limiting its practical utility in real-world scenarios.

Moreover, the construction of the dataset and the nature of the evaluation metric also have a significant impact on the assessment of the performance of the models. Although the "Perfect Predictions" metric offers a clear evaluation scheme, it does not consider the inherent ambiguity of coding solutions. In the real world of code repair, there are often multiple correct ways to solve a problem, but this metric only validates one specific version of the solution. This type of approach may underestimate the true capabilities of the model, as it may miss perfectly acceptable solutions that do not exactly match the reference output.

Future research should focus on developing more nuanced evaluation metrics to overcome the limitations of the "Perfect Predictions" approach. While some studies have employed BLEU and CodeBLEU scores (Zhou, Kim, et al., 2024), these metrics also have limitations in the context of code repair. They primarily assess textual similarity rather than functional correctness or security improvement, which are crucial in vulnerability fixing. Instead, future metrics could consider functionally equivalent solutions and assess the quality of repairs beyond exact text matches. Additionally, integrating static code analyzers with our models could enhance vulnerability detection and provide more context for accurate fixes. This combined approach may lead to more comprehensive and reliable automated repair systems.

Another promising direction is the implementation of an intelligent patch selection method. This could include the development of a ranking system to identify the most probable fix among multiple patches generated, potentially improving the overall effectiveness of the repair process. In addition, exploring the generation of synthetic data could significantly improve model training and evaluation. By creating diverse, artificially generated vulnerability samples, we could address the current limitations in dataset quality and availability (Zhou, Cao, et al., 2024). This approach would not only increase the volume of training data, but also ensure a more representative range of vulnerability types and coding patterns. Synthetic data generation could help create more robust models capable of handling a wider range of real-world vulnerabilities, thus improving their practical applicability in diverse software environments.

In conclusion, our research represents a significant advancement in the field of automated source code vulnerability repair by presenting a comprehensive approach that combines innovative representation methods, rigorous dataset management, and the strategic use of advanced LLMs. We have not only shown the efficacy of our method in contrast to established approaches, but also highlighted important aspects for further improvement, particularly in dataset creation and performance evaluation metrics. Our findings emphasize the significance of precise data representation and the need for adaptable and nuanced evaluation techniques to truly move forward in the field. This research provides a foundation for developing automated code repair solutions that are more reliable, efficient, and versatile. Ultimately, this will improve the security and integrity of software systems in the rapidly changing digital landscape.

## Acknowledgements

The authors want to thank the support by the research project Contract No. TASE-UAH-23.032. Artificial Intelligence applied to code repair after code static analysis verification.

## References

Ahmad, W., Chakraborty, S., Ray, B., & Chang, K.-W. (2021). Unified Pre-training for Program Understanding and Generation. In K. Toutanova, A. Rumshisky, L. Zettlemoyer, D. Hakkani-Tur, I. Beltagy, S. Bethard, R. Cotterell, T. Chakraborty, & Y. Zhou (Eds.), *Proceedings of the 2021 Conference of the North American Chapter of the Association for Computational Linguistics: Human Language Technologies* (pp. 2655–2668). Association for Computational Linguistics. https://doi.org/10.18653/v1/2021.naacl-main.211

Berabi, B., He, J., Raychev, V., & Vechev, M. (2021). TFix: Learning to Fix Coding Errors with a Text-to-Text Transformer. In M. Meila & T. Zhang (Eds.), *Proceedings of the 38th International Conference on Machine Learning* (Vol. 139, pp. 780–791). PMLR. https://proceedings.mlr.press/v139/berabi21a.html

Bhandari, G., Naseer, A., & Moonen, L. (2021). CVEfixes: Automated Collection of Vulnerabilities and Their Fixes from Open-Source Software. *Proceedings of the 17th International Conference on Predictive Models and Data Analytics in Software Engineering*, 30–39. https://doi.org/10.1145/3475960.3475985

Brown, T., Mann, B., Ryder, N., Subbiah, M., Kaplan, J. D., Dhariwal, P., Neelakantan, A., Shyam, P., Sastry, G., Askell, A., Agarwal, S., Herbert-Voss, A., Krueger, G., Henighan, T., Child, R., Ramesh, A., Ziegler, D., Wu, J., Winter, C., … Amodei, D. (2020). Language Models are Few-Shot Learners. In H. Larochelle, M. Ranzato, R. Hadsell, M. F. Balcan, & H. Lin (Eds.), *Advances in Neural Information Processing Systems* (Vol. 33, pp. 1877–1901). Curran Associates, Inc. https://proceedings.neurips.cc/paper_files/paper/2020/file/1457c0d6bfcb4967418bfb8ac142f64a-Paper.pdf

Chen, M., Tworek, J., Jun, H., Yuan, Q., Pinto, H. P. de O., Kaplan, J., Edwards, H., Burda, Y., Joseph, N., Brockman, G., Ray, A., Puri, R., Krueger, G., Petrov, M., Khlaaf, H., Sastry, G., Mishkin, P., Chan, B., Gray, S., … Zaremba, W. (2021). *Evaluating Large Language Models Trained on Code* (arXiv:2107.03374). arXiv. https://doi.org/10.48550/arXiv.2107.03374

Chen, Z., Kommrusch, S., & Monperrus, M. (2022). Neural transfer learning for repairing security vulnerabilities in c code. *IEEE Transactions on Software Engineering*, *49*(1), 147–165.

Chowdhery, A., Narang, S., Devlin, J., Bosma, M., Mishra, G., Roberts, A., Barham, P., Chung, H. W., Sutton, C., Gehrmann, S., Schuh, P., Shi, K., Tsvyashchenko, S., Maynez, J., Rao, A., Barnes, P., Tay, Y., Shazeer, N., Prabhakaran, V., … Fiedel, N. (2023). PaLM: Scaling Language Modeling with Pathways. *Journal of Machine Learning Research*, *24*(240), 1–113.

Dettmers, T., Pagnoni, A., Holtzman, A., & Zettlemoyer, L. (2023). *QLoRA: Efficient Finetuning of Quantized LLMs* (arXiv:2305.14314). arXiv. https://doi.org/10.48550/arXiv.2305.14314

Dettmers, T., & Zettlemoyer, L. (2023). *The case for 4-bit precision: K-bit Inference Scaling Laws* (arXiv:2212.09720). arXiv. https://doi.org/10.48550/arXiv.2212.09720

Devlin, J., Chang, M.-W., Lee, K., & Toutanova, K. (2019). *BERT: Pre-training of Deep Bidirectional Transformers for Language Understanding* (arXiv:1810.04805). arXiv. https://doi.org/10.48550/arXiv.1810.04805

Du, N., Huang, Y., Dai, A. M., Tong, S., Lepikhin, D., Xu, Y., Krikun, M., Zhou, Y., Yu, A. W., Firat, O., Zoph, B., Fedus, L., Bosma, M. P., Zhou, Z., Wang, T., Wang, E., Webster, K., Pellat, M., Robinson, K., … Cui, C. (2022). GLaM: Efficient Scaling of Language Models with Mixture-of-Experts. In K. Chaudhuri, S. Jegelka, L. Song, C. Szepesvari, G. Niu, & S. Sabato (Eds.), *Proceedings of the 39th International Conference on Machine Learning* (Vol. 162, pp. 5547–5569). PMLR. https://proceedings.mlr.press/v162/du22c.html

Durieux, T., & Monperrus, M. (2016). DynaMoth: Dynamic code synthesis for automatic program repair. *Proceedings of the 11th International Workshop on Automation of Software Test*, 85–91. https://doi.org/10.1145/2896921.2896931

Fan, J., Li, Y., Wang, S., & Nguyen, T. N. (2020). A C/C++ Code Vulnerability Dataset with Code Changes and CVE Summaries. *Proceedings of the 17th International Conference on Mining Software Repositories*, 508–512. https://doi.org/10.1145/3379597.3387501

Farahbod, K., Shayo, C., & Varzandeh, J. (2020). Cybersecurity indices and cybercrime annual loss and economic impacts. *Journal of Business and Behavioral Sciences*, *32*(1), 63–71.

Feng, Z., Guo, D., Tang, D., Duan, N., Feng, X., Gong, M., Shou, L., Qin, B., Liu, T., Jiang, D., & Zhou, M. (2020). *CodeBERT: A Pre-Trained Model for Programming and Natural Languages* (arXiv:2002.08155). arXiv. https://doi.org/10.48550/arXiv.2002.08155

Frantar, E., Ashkboos, S., Hoefler, T., & Alistarh, D. (2023). *GPTQ: Accurate Post-Training Quantization for Generative Pre-trained Transformers* (arXiv:2210.17323). arXiv. https://doi.org/10.48550/arXiv.2210.17323

Fried, D., Aghajanyan, A., Lin, J., Wang, S., Wallace, E., Shi, F., Zhong, R., Yih, W., Zettlemoyer, L., & Lewis, M. (2023). *InCoder: A Generative Model for Code Infilling and Synthesis* (arXiv:2204.05999). arXiv. https://doi.org/10.48550/arXiv.2204.05999

Fu, M., Tantithamthavorn, C., Le, T., Nguyen, V., & Phung, D. (2022). VulRepair: A T5-based automated software vulnerability repair. *Proceedings of the 30th ACM Joint European Software Engineering Conference and Symposium on the Foundations of Software Engineering*, 935–947. https://doi.org/10.1145/3540250.3549098

Gao, X., Noller, Y., & Roychoudhury, A. (2022). *Program Repair* (arXiv:2211.12787). arXiv. https://doi.org/10.48550/arXiv.2211.12787

Guo, D., Lu, S., Duan, N., Wang, Y., Zhou, M., & Yin, J. (2022). *UniXcoder: Unified Cross-Modal Pre-training for Code Representation* (arXiv:2203.03850). arXiv. https://doi.org/10.48550/arXiv.2203.03850

Guo, D., Ren, S., Lu, S., Feng, Z., Tang, D., Liu, S., Zhou, L., Duan, N., Svyatkovskiy, A., Fu, S., Tufano, M., Deng, S. K., Clement, C., Drain, D., Sundaresan, N., Yin, J., Jiang, D., & Zhou, M. (2021). *GraphCodeBERT: Pre-training Code Representations with Data Flow* (arXiv:2009.08366). arXiv. https://doi.org/10.48550/arXiv.2009.08366

Harer, J., Ozdemir, O., Lazovich, T., Reale, C., Russell, R., Kim, L., & chin, peter. (2018). Learning to Repair Software Vulnerabilities with Generative Adversarial Networks. *Advances in Neural Information Processing Systems*, *31*. https://proceedings.neurips.cc/paper/2018/hash/68abef8ee1ac9b664a90b0bbaff4f770-Abstract.html

Hoffmann, J., Borgeaud, S., Mensch, A., Buchatskaya, E., Cai, T., Rutherford, E., Casas, D. de L., Hendricks, L. A., Welbl, J., Clark, A., Hennigan, T., Noland, E., Millican, K., Driessche, G. van den, Damoc, B., Guy, A., Osindero, S., Simonyan, K., Elsen, E., … Sifre, L. (2022). *Training Compute-Optimal Large Language Models* (arXiv:2203.15556). arXiv. https://doi.org/10.48550/arXiv.2203.15556

Howard, J., & Ruder, S. (2018). *Universal Language Model Fine-tuning for Text Classification* (arXiv:1801.06146). arXiv. https://doi.org/10.48550/arXiv.1801.06146

Hu, E. J., Shen, Y., Wallis, P., Allen-Zhu, Z., Li, Y., Wang, S., Wang, L., & Chen, W. (2021). *LoRA: Low-Rank Adaptation of Large Language Models* (arXiv:2106.09685). arXiv. https://doi.org/10.48550/arXiv.2106.09685

Hu, Y., Ahmed, U. Z., Mechtaev, S., Leong, B., & Roychoudhury, A. (2019). Re-Factoring Based Program Repair Applied to Programming Assignments. *2019 34th IEEE/ACM*

*International Conference on Automated Software Engineering (ASE)*, 388–398. https://doi.org/10.1109/ASE.2019.00044

Huang, K., Meng, X., Zhang, J., Liu, Y., Wang, W., Li, S., & Zhang, Y. (2023). An Empirical Study on Fine-Tuning Large Language Models of Code for Automated Program Repair. *2023 38th IEEE/ACM International Conference on Automated Software Engineering (ASE)*, 1162–1174. https://doi.org/10.1109/ASE56229.2023.00181

Huang, Z., Lie, D., Tan, G., & Jaeger, T. (2019). Using Safety Properties to Generate Vulnerability Patches. *2019 IEEE Symposium on Security and Privacy (SP)*, 539–554. https://doi.org/10.1109/SP.2019.00071

Ishizue, R., Sakamoto, K., Washizaki, H., & Fukazawa, Y. (2024). Improved Program Repair Methods using Refactoring with GPT Models. *Proceedings of the 55th ACM Technical Symposium on Computer Science Education V. 1*, 569–575. https://doi.org/10.1145/3626252.3630875

Islam, N. T., Khoury, J., Seong, A., Karkevandi, M. B., Parra, G. D. L. T., Bou-Harb, E., & Najafirad, P. (2024). *LLM-Powered Code Vulnerability Repair with Reinforcement Learning and Semantic Reward* (arXiv:2401.03374). arXiv. https://doi.org/10.48550/arXiv.2401.03374

Jain, N., Chiang, P., Wen, Y., Kirchenbauer, J., Chu, H.-M., Somepalli, G., Bartoldson, B. R., Kailkhura, B., Schwarzschild, A., Saha, A., Goldblum, M., Geiping, J., & Goldstein, T. (2023). *NEFTune: Noisy Embeddings Improve Instruction Finetuning* (arXiv:2310.05914). arXiv. https://doi.org/10.48550/arXiv.2310.05914

Ji, T., Wu, Y., Wang, C., Zhang, X., & Wang, Z. (2018). The Coming Era of AlphaHacking?: A Survey of Automatic Software Vulnerability Detection, Exploitation and Patching Techniques. *2018 IEEE Third International Conference on Data Science in Cyberspace (DSC)*, 53–60. https://doi.org/10.1109/DSC.2018.00017

Jiang, A. Q., Sablayrolles, A., Mensch, A., Bamford, C., Chaplot, D. S., Casas, D. de las, Bressand, F., Lengyel, G., Lample, G., Saulnier, L., Lavaud, L. R., Lachaux, M.-A., Stock, P., Scao, T. L., Lavril, T., Wang, T., Lacroix, T., & Sayed, W. E. (2023). *Mistral 7B* (arXiv:2310.06825). arXiv. https://doi.org/10.48550/arXiv.2310.06825

Jiang, J., Xiong, Y., Zhang, H., Gao, Q., & Chen, X. (2018). Shaping program repair space with existing patches and similar code. *Proceedings of the 27th ACM SIGSOFT International Symposium on Software Testing and Analysis*, 298–309. https://doi.org/10.1145/3213846.3213871

Kaplan, J., McCandlish, S., Henighan, T., Brown, T. B., Chess, B., Child, R., Gray, S., Radford, A., Wu, J., & Amodei, D. (2020). *Scaling Laws for Neural Language Models* (arXiv:2001.08361). arXiv. https://doi.org/10.48550/arXiv.2001.08361

Koyuncu, A., Liu, K., Bissyandé, T. F., Kim, D., Klein, J., Monperrus, M., & Le Traon, Y. (2020). FixMiner: Mining relevant fix patterns for automated program repair. *Empirical Software Engineering*, *25*(3), 1980–2024. https://doi.org/10.1007/s10664-019-09780-z

Kwon, W., Li, Z., Zhuang, S., Sheng, Y., Zheng, L., Yu, C. H., Gonzalez, J. E., Zhang, H., & Stoica, I. (2023). *Efficient Memory Management for Large Language Model Serving with PagedAttention* (arXiv:2309.06180). arXiv. https://doi.org/10.48550/arXiv.2309.06180

Le Goues, C., Nguyen, T., Forrest, S., & Weimer, W. (2012). GenProg: A Generic Method for Automatic Software Repair. *IEEE Transactions on Software Engineering*, *38*(1), 54–72. IEEE Transactions on Software Engineering. https://doi.org/10.1109/TSE.2011.104

Lin, J., Tang, J., Tang, H., Yang, S., Dang, X., Gan, C., & Han, S. (2023). *AWQ: Activation-aware Weight Quantization for LLM Compression and Acceleration* (arXiv:2306.00978). arXiv. https://doi.org/10.48550/arXiv.2306.00978

Lin, Z., Jiang, X., Xu, D., Mao, B., & Xie, L. (2007). AutoPaG: Towards automated software patch generation with source code root cause identification and repair. *Proceedings of the 2nd ACM Symposium on Information, Computer and Communications Security*, 329–340. https://doi.org/10.1145/1229285.1267001

Liu, K., Koyuncu, A., Kim, D., & Bissyandé, T. F. (2019). TBar: Revisiting template-based automated program repair. *Proceedings of the 28th ACM SIGSOFT International Symposium on Software Testing and Analysis*, 31–42. https://doi.org/10.1145/3293882.3330577

Lomio, F., Iannone, E., De Lucia, A., Palomba, F., & Lenarduzzi, V. (2022). Just-in-time software vulnerability detection: Are we there yet? *Journal of Systems and Software*, *188*, 111283. https://doi.org/10.1016/j.jss.2022.111283

Loshchilov, I., & Hutter, F. (2019). *Decoupled Weight Decay Regularization* (arXiv:1711.05101). arXiv. https://doi.org/10.48550/arXiv.1711.05101

Lu, S., Guo, D., Ren, S., Huang, J., Svyatkovskiy, A., Blanco, A., Clement, C., Drain, D., Jiang, D., Tang, D., Li, G., Zhou, L., Shou, L., Zhou, L., Tufano, M., Gong, M., Zhou, M., Duan, N., Sundaresan, N., … Liu, S. (2021). *CodeXGLUE: A Machine Learning Benchmark Dataset for Code Understanding and Generation* (arXiv:2102.04664). arXiv. https://doi.org/10.48550/arXiv.2102.04664

Ma, S., Thung, F., Lo, D., Sun, C., & Deng, R. H. (2017). VuRLE: Automatic Vulnerability Detection and Repair by Learning from Examples. In S. N. Foley, D. Gollmann, & E. Snekkenes (Eds.), *Computer Security – ESORICS 2017* (pp. 229–246). Springer International Publishing. https://doi.org/10.1007/978-3-319-66399-9_13

Martinez, M., & Monperrus, M. (2016). ASTOR: A program repair library for Java (demo). *Proceedings of the 25th International Symposium on Software Testing and Analysis*, 441–444. https://doi.org/10.1145/2931037.2948705

Martinez, M., & Monperrus, M. (2018). Ultra-Large Repair Search Space with Automatically Mined Templates: The Cardumen Mode of Astor. In T. E. Colanzi & P. McMinn (Eds.), *Search-Based Software Engineering* (pp. 65–86). Springer International Publishing. https://doi.org/10.1007/978-3-319-99241-9_3

Mechtaev, S., Yi, J., & Roychoudhury, A. (2016). Angelix: Scalable multiline program patch synthesis via symbolic analysis. *Proceedings of the 38th International Conference on Software Engineering*, 691–701. https://doi.org/10.1145/2884781.2884807

Nijkamp, E., Hayashi, H., Xiong, C., Savarese, S., & Zhou, Y. (2023). *CodeGen2: Lessons for Training LLMs on Programming and Natural Languages* (arXiv:2305.02309). arXiv. https://doi.org/10.48550/arXiv.2305.02309

Nijkamp, E., Pang, B., Hayashi, H., Tu, L., Wang, H., Zhou, Y., Savarese, S., & Xiong, C. (2023). *CodeGen: An Open Large Language Model for Code with Multi-Turn Program Synthesis* (arXiv:2203.13474). arXiv. https://doi.org/10.48550/arXiv.2203.13474

*Octoverse 2022: The state of open source*. (2022). The State of the Octoverse. https://octoverse.github.com/

O'Driscoll, A. (2022). Cyber security vulnerability statistics and facts of 2022. *Comparitech*.

OpenAI. (2023). *GPT-4 Technical Report*.

Ouyang, L., Wu, J., Jiang, X., Almeida, D., Wainwright, C., Mishkin, P., Zhang, C., Agarwal, S., Slama, K., Ray, A., Schulman, J., Hilton, J., Kelton, F., Miller, L., Simens, M., Askell, A., Welinder, P., Christiano, P. F., Leike, J., & Lowe, R. (2022). Training language models to follow instructions with human feedback. In S. Koyejo, S. Mohamed, A. Agarwal, D. Belgrave, K. Cho, & A. Oh (Eds.), *Advances in Neural Information Processing Systems* (Vol. 35, pp. 27730–27744). Curran Associates, Inc. https://proceedings.neurips.cc/paper_files/paper/2022/file/b1efde53be364a73914f58805a001731-Paper-Conference.pdf

Radford, A., Narasimhan, K., Salimans, T., & Sutskever, I. (2018). *Improving language understanding by generative pre-training*.

Radford, A., Wu, J., Child, R., Luan, D., Amodei, D., Sutskever, I., & others. (2019). Language models are unsupervised multitask learners. *OpenAI Blog*, *1*(8), 9.

Rae, J. W., Borgeaud, S., Cai, T., Millican, K., Hoffmann, J., Song, F., Aslanides, J., Henderson, S., Ring, R., Young, S., Rutherford, E., Hennigan, T., Menick, J., Cassirer, A., Powell, R., Driessche, G. van den, Hendricks, L. A., Rauh, M., Huang, P.-S., … Irving, G. (2022). *Scaling Language Models: Methods, Analysis & Insights from Training Gopher* (arXiv:2112.11446). arXiv. https://doi.org/10.48550/arXiv.2112.11446

Raffel, C., Shazeer, N., Roberts, A., Lee, K., Narang, S., Matena, M., Zhou, Y., Li, W., & Liu, P. J. (2020). Exploring the Limits of Transfer Learning with a Unified Text-to-Text Transformer. *Journal of Machine Learning Research*, *21*(140), 1–67.

Rozière, B., Gehring, J., Gloeckle, F., Sootla, S., Gat, I., Tan, X. E., Adi, Y., Liu, J., Remez, T., Rapin, J., Kozhevnikov, A., Evtimov, I., Bitton, J., Bhatt, M., Ferrer, C. C., Grattafiori, A., Xiong, W., Défossez, A., Copet, J., … Synnaeve, G. (2023). *Code Llama: Open Foundation Models for Code* (arXiv:2308.12950). arXiv. https://doi.org/10.48550/arXiv.2308.12950

Shariffdeen, R., Noller, Y., Grunske, L., & Roychoudhury, A. (2021). Concolic program repair. *Proceedings of the 42nd ACM SIGPLAN International Conference on Programming Language Design and Implementation*, 390–405. https://doi.org/10.1145/3453483.3454051

Sidiroglou, S., & Keromytis, A. D. (2005). Countering network worms through automatic patch generation. *IEEE Security & Privacy*, *3*(6), 41–49.

Smith, S., Patwary, M., Norick, B., LeGresley, P., Rajbhandari, S., Casper, J., Liu, Z., Prabhumoye, S., Zerveas, G., Korthikanti, V., Zhang, E., Child, R., Aminabadi, R. Y., Bernauer, J., Song, X., Shoeybi, M., He, Y., Houston, M., Tiwary, S., & Catanzaro, B. (2022). *Using DeepSpeed and Megatron to Train Megatron-Turing NLG 530B, A Large-Scale Generative Language Model* (arXiv:2201.11990). arXiv. https://doi.org/10.48550/arXiv.2201.11990

Tay, Y., Dehghani, M., Tran, V. Q., Garcia, X., Wei, J., Wang, X., Chung, H. W., Shakeri, S., Bahri, D., Schuster, T., Zheng, H. S., Zhou, D., Houlsby, N., & Metzler, D. (2023). *UL2: Unifying Language Learning Paradigms* (arXiv:2205.05131). arXiv. https://doi.org/10.48550/arXiv.2205.05131

Thoppilan, R., De Freitas, D., Hall, J., Shazeer, N., Kulshreshtha, A., Cheng, H.-T., Jin, A., Bos, T., Baker, L., Du, Y., Li, Y., Lee, H., Zheng, H. S., Ghafouri, A., Menegali, M., Huang, Y., Krikun,

M., Lepikhin, D., Qin, J., … Le, Q. (2022). *LaMDA: Language Models for Dialog Applications* (arXiv:2201.08239). arXiv. https://doi.org/10.48550/arXiv.2201.08239

Touvron, H., Martin, L., Stone, K., Albert, P., Almahairi, A., Babaei, Y., Bashlykov, N., Batra, S., Bhargava, P., Bhosale, S., Bikel, D., Blecher, L., Ferrer, C. C., Chen, M., Cucurull, G., Esiobu, D., Fernandes, J., Fu, J., Fu, W., … Scialom, T. (2023). *Llama 2: Open Foundation and Fine-Tuned Chat Models* (arXiv:2307.09288). arXiv. https://doi.org/10.48550/arXiv.2307.09288

Tunstall, L., von Werra, L., & Wolf, T. (2022). *Natural Language Processing with Transformers*. O'Reilly Media, Inc.

Vaswani, A., Shazeer, N., Parmar, N., Uszkoreit, J., Jones, L., Gomez, A. N., Kaiser, Ł., & Polosukhin, I. (2017). Attention is All you Need. In I. Guyon, U. V. Luxburg, S. Bengio, H. Wallach, R. Fergus, S. Vishwanathan, & R. Garnett (Eds.), *Advances in Neural Information Processing Systems* (Vol. 30). Curran Associates, Inc. https://proceedings.neurips.cc/paper_files/paper/2017/file/3f5ee243547dee91fbd053c1c4a845aa-Paper.pdf

Wang, Y., Wang, W., Joty, S., & Hoi, S. C. H. (2021). *CodeT5: Identifier-aware Unified Pre-trained Encoder-Decoder Models for Code Understanding and Generation* (arXiv:2109.00859). arXiv. https://doi.org/10.48550/arXiv.2109.00859

Wu, Y., Jiang, N., Pham, H. V., Lutellier, T., Davis, J., Tan, L., Babkin, P., & Shah, S. (2023). How Effective Are Neural Networks for Fixing Security Vulnerabilities. *Proceedings of the 32nd ACM SIGSOFT International Symposium on Software Testing and Analysis*, 1282–1294. https://doi.org/10.1145/3597926.3598135

Xia, C. S., Wei, Y., & Zhang, L. (2023). Automated Program Repair in the Era of Large Pre-trained Language Models. *2023 IEEE/ACM 45th International Conference on Software Engineering (ICSE)*, 1482–1494. https://doi.org/10.1109/ICSE48619.2023.00129

Xuan, J., Martinez, M., DeMarco, F., Clément, M., Marcote, S. L., Durieux, T., Le Berre, D., & Monperrus, M. (2017). Nopol: Automatic Repair of Conditional Statement Bugs in Java Programs. *IEEE Transactions on Software Engineering*, *43*(1), 34–55. IEEE Transactions on Software Engineering. https://doi.org/10.1109/TSE.2016.2560811

Yuan, Y., & Banzhaf, W. (2020). ARJA: Automated Repair of Java Programs via Multi-Objective Genetic Programming. *IEEE Transactions on Software Engineering*, *46*(10), 1040–1067. IEEE Transactions on Software Engineering. https://doi.org/10.1109/TSE.2018.2874648

Zhang, Q., Fang, C., Ma, Y., Sun, W., & Chen, Z. (2023). A Survey of Learning-based Automated Program Repair. *ACM Trans. Softw. Eng. Methodol.*, *33*(2). https://doi.org/10.1145/3631974

Zhang, Q., Zhao, Y., Sun, W., Fang, C., Wang, Z., & Zhang, L. (2022). *Program Repair: Automated vs. Manual* (arXiv:2203.05166). arXiv. https://doi.org/10.48550/arXiv.2203.05166

Zhou, X., Cao, S., Sun, X., & Lo, D. (2024). *Large Language Model for Vulnerability Detection and Repair: Literature Review and the Road Ahead* (arXiv:2404.02525). arXiv. https://doi.org/10.48550/arXiv.2404.02525

Zhou, X., Kim, K., Xu, B., Han, D., & Lo, D. (2024). *Multi-LLM Collaboration + Data-Centric Innovation = 2x Better Vulnerability Repair* (arXiv:2401.15459). arXiv. https://doi.org/10.48550/arXiv.2401.15459

Zirak, A., & Hemmati, H. (2024). Improving Automated Program Repair with Domain Adaptation. *ACM Trans. Softw. Eng. Methodol.*, *33*(3), 65:1-65:43. https://doi.org/10.1145/3631972